\documentclass[a4paper]{revtex4}

\usepackage[english]{babel}
\usepackage{amsmath}
\usepackage{amssymb}
\usepackage{amsbsy}
\usepackage{amstext}
\usepackage{graphicx}
\usepackage{subfigure}
\usepackage{pst-node}
\usepackage{verbatim}
\usepackage{color}
\usepackage{ulem}   %% for markup
\newcommand{\be}{\begin{eqnarray}}
\newcommand{\ee}{\end{eqnarray}}
\newcommand{\bdm}{\begin{displaymath}}
\newcommand{\edm}{\end{displaymath}}
\newcommand{\ds}{\displaystyle}
\newcommand{\ba}{\begin{array}}
\newcommand{\ea}{\end{array}}
\newcommand{\pa}[1]{\left(#1\right)}
\newcommand{\paq}[1]{\left[#1\right]}
\newcommand{\omm}{{\omega_m}}

\newcommand{\omrd}{{\omega_{rd}}}
\begin{document}

\title{Complete phenomenological gravitational waveforms from spinning coalescing binaries}

\author{R. Sturani$^{(1,2)}$, S. Fischetti$^{(3)}$, L. Cadonati$^{(3)}$, G. M. 
Guidi$^{(1,2)}$, J. Healy$^{(4)}$, D. Shoemaker$^{(4)}$, A. Vicer\'e$^{(1,2)}$}

\address{(1) Dipartimento di Matematica, Fisica e Informatica, Universit\`a 
  degli Studi di Urbino `Carlo Bo', I-61029 Urbino, Italy\\
  (2) INFN Sezione di Firenze, I-50019 Sesto Fiorentino, Italy\\
  (3) Physics Department, University of Massachusetts, Amherst MA 01003,  USA\\
  (4) Center for Relativistic Astrophysics, Georgia Tech, Atlanta, GA 30332, USA}

\email{riccardo.sturani@uniurb.it}

\begin{abstract}
  The quest for gravitational waves from coalescing binaries is customarily 
  performed by the LIGO-Virgo collaboration via 
  matched filtering, which requires a detailed knowledge of the signal. 
  Complete analytical coalescence waveforms are currently available only for 
  the non-precessing binary systems. In this paper we introduce
  complete phenomenological waveforms for
  the dominant quadrupolar mode of generically spinning systems.
  These waveforms are constructed by bridging the gap between the analytically 
  known inspiral phase, described by spin Taylor (T4) approximants in the 
  restricted  waveform approximation, and the ring-down phase through a 
  phenomenological intermediate phase, calibrated by comparison with 
  specific, numerically generated waveforms, describing equal mass systems
  with dimension-less spin magnitudes equal to 0.6.
  The overlap integral between numerical and phenomenological waveforms ranges 
  between 0.95 and 0.99.
\end{abstract}


\maketitle

\section{Introduction}

The Laser Interferometer Gravitational-wave Observatory (LIGO) and Virgo
constitute a network of kilometer-scale interferometers for the detection of 
gravitational waves. 
The initial detector configuration has successfully acquired data at design 
sensitivity; substantial upgrades over the next few years, to Advanced LIGO and
Advanced Virgo, are expected to yield the detection of gravitational waves from
the coalescence of compact binary systems of neutron stars or black holes.

The coalescence of binary systems is usually described in terms of three
distinct phases: the \emph{inspiral}, the \emph{merger} and the 
\emph{ring-down}. The inspiral phase allows for an accurate analytical 
description via the so-called Post-Newtonian (PN) expansion, see for 
instance \cite{Blanchet:2006zz} for a review. 
The ring-down also admits a perturbative analytical model, as it describes the 
damped oscillations of the single object resulting from the binary coalescence,
as a superposition of black-hole quasi normal modes \cite{Teukolsky:1973ha}. 
The merger phase is however fully non-perturbative
and for generic systems it has not been described analytically but
rather by numerical simulations. During the last five years numerical 
relativity has made tremendous progress in describing the full coalescence
of a binary system 
beginning with \cite{Pretorius:2005gq,Campanelli:2005dd, Baker:2005vv} and more
recently~\cite{Herrmann:2007ex,Tichy:2008du,Scheel:2008rj,Baker:2008mj,Ajith:2009bn,Lousto:2009ka,Pollney:2007ss}, 
see~\cite{Hannam:2009rd,Husa:2007zz,Pretorius:2007nq} for reviews
and it can now produce waveforms for generic spin orientations, with moderate 
spin magnitude ($\lesssim 0.9$) and mass ratios ($\lesssim 10:1$).

Matched filtering  is typically used in LIGO-Virgo data analysis in order to 
uncover weak signals buried into noise \cite{:2010uf}.
This technique requires a large number (tens of thousands) of templates to be 
checked against real data. Due to the computational cost of numerical 
simulations, this is only possible with an analytical
%thus requiring for practical reasons an analytic 
knowledge of the templates.
Analytical methods are currently available to reproduce the complete
waveform emitted by non-precessing coalescing binaries. This has been achieved 
in the Effective One Body (EOB) construction 
\cite{Buonanno:1998gg,Buonanno:2006ui,Damour:2009kr}
for non-spinning systems, and in the 
EOB-spin waveforms \cite{Pan:2009wj}, an extension of the EOB method to the 
case of binaries with non-vanishing spins aligned with the orbital angular 
momentum. Both spinning and non spinning EOB require a comparison
with numerically generated waveforms in order to calibrate some free 
parameters of the model.
Another method for generating analytical waveforms for
spinning non-precessing binaries is  by joining 
PN-generated inspiral and numerical waveforms \cite{Ajith:2009bn,Ajith:2007qp}
to construct \emph{phenomenological} waveforms.

In this paper we present a new family of analytical 
waveforms from the coalescence of generic spinning binaries.

%The waveforms \xout{subject of the present study} have been constructed by 
%joining the perturbative PN description to the ring-down phase by a phenomenological 
%phase allowing to interpolate between the inspiral and the  ring-down. 
The waveforms are obtained by interpolating between the perturbative PN 
inspiral description and the ring-down, both admitting physically motivated
analytic models, with a phenomenological phase over the merger portion of the 
signal.
Since in the matched-filtering search an accurate determination of 
the gravitational phase of the signal is crucial, more important than an 
accurate amplitude determination, we focus on the phenomenological 
parametrization of the gravitational wave phase, and give lower priority to the
amplitude.\\
%the phenomenological parametrization of the waveform
%on the phase modeling, and keep the amplitude at lowest order.\\
Spinning waveforms depend on several parameters (masses, spin
components of the binary constituents, angles defining the orientation
of the source with respect to the observer). For this reason it is not practical
to use these waveforms to build template banks, but we envision their usefulness
as injection waveforms in testing existing LIGO-Virgo data analysis pipelines, 
which are currently based on non-spinning templates.
%The dependence of spinning waveforms on several parameters (masses plus spin
%components of the binary constituents) will make impractical their use as 
%template banks, but they could provide useful tools for testing the 
%present data analysis pipeline, based on non-spinning templates.

The paper is organized as follows.
In sec.~\ref{se:method} the method for the analytical waveform construction and 
the numerical simulations are illustrated. In sec.~\ref{se:result}
the results are presented, in the form of comparison between analytically and
numerically generated waveforms. We restricted the analysis to the main 
quadrupolar mode ($m=2$), 
%but it is straightforward in this setup 
but this setup can be extended to include other modes.
In sec.~\ref{se:concl} the conclusions that can be drawn from the present 
work are reported.

\section{The method}
\label{se:method}

The goal of the present work is to produce the complete analytical waveforms 
generated by the coalescence of spinning binary systems.
The numerical waveforms used to construct and calibrate our analytical model
all describe equal mass binary systems ($m_1=m_2$), with spin magnitudes 
$|{\mathbf {S_1}}|=|\mathbf{S_2}|=0.6\, m_1^2$ and $\mathbf{S_2}$ orthogonal 
to the initial orbital angular momentum (where $\mathbf{S_{1,2}}$
denote the binary constituent spin vectors and we posit $G_N=c=1$).

The description of the dynamics adopted here models the inspiral phase via 
the standard TaylorT4 PN formulas, see \cite{Buonanno:2009zt} for definition 
and comparison of different PN approximants in the spin-less case. 
In the non-spinning case the waveform is fully parametrized by amplitude and 
the orbital phase, but in the spinning case the effect of the precession of the 
orbital plane should also be accounted for, see \cite{Racine:2008kj} spin and 
angular momentum evolution equations.\\
It is convenient to define an \emph{orbital} phase 
$\phi=\int \omega_{orb}\,dt$ whose evolution is given by
\begin{equation}
\frac{d\phi}{dt}=\frac{v^3}M\,, \qquad  \frac{dv}{dt}=-\frac{F(v)}{dE/dv}\,,
\end{equation}
where $M\equiv m_1+m_2$ is the total mass of the binary system, $F(v)$ and 
$E(v)$ are respectively the flux emitted and the energy of a circular orbit
with angular frequency $\omega_{orb}=v^3/M$. The main gravitational wave 
frequency $f_{GW}$ is related to $\omega_{orb}$ via $f_{GW}=\omega_{orb}/\pi$.\\
In order to determine the actual 
waveform also the spins and the orbital angular momentum have to be dynamically
evolved, see e.g. \cite{Arun:2008kb}. In particular by parametrizing the orbital
angular momentum unit vector $\hat{\mathbf L}$ as 
${\mathbf{\hat L}=(\sin\iota\cos\alpha,\sin\iota\sin\alpha,\cos\iota)}$ 
%(in the frame where the total angular momentum is parallel to the $\hat z$ axis), 
it is convenient to introduce the \emph{carrier} phase $\Psi$ given by
\be
\frac {d\Psi}{dt}=\omega_{orb}-\cos\iota\,\frac{d\alpha}{dt}\,.
\ee
Numerically generated waveforms are usually decomposed in spherical harmonics,
in particular the five quadrupolar modes ($l=2$) are the only non-vanishing
at the lowest order in $v$, and the $l=2,m=\pm 2$ mode are the dominant ones. 
The actual shape of the $l=2,m=2$ mode, the only one which will be used here
for comparison with numerical relativity results, is
given by the following \cite{Arun:2008kb}
\renewcommand{\arraystretch}{1.4}
\be
\label{eq:h22}
\ba{rcl}
h^{(insp)}_{2,2}&=&\ds -2\frac{\nu Mv^2}{d}\sqrt{\frac{16\pi}{5}}
\paq{\cos^4(\iota/2)\cos\pa{2(\Psi+\alpha)}+
\sin^4(\iota/2)\cos\pa{2(\Psi-\alpha)}+O(v)}\,,\\
%h^{(insp)}_{2,-2}&=&\ds -2\frac{\nu Mv^2}{d}\sqrt{\frac{16\pi}{5}}
%\paq{-\cos^4(\iota/2)\sin\pa{2(\Psi+\alpha)}+
%\sin^4(\iota/2)\sin\pa{2(\Psi-\alpha)}+O(v)}\,,
%h_{2,1}=+O(v)\,,
%h_{2,0}=+O(v)\,.
\ea
\ee
\renewcommand{\arraystretch}{1}
where $\nu\equiv m_1m_2/M^2$ is the symmetric mass ratio, $d$ the 
source-observer distance and this formula refers to the inspiral waveform only.
\\
The functions $F(v)$ and $E(v)$, necessary to determine the orbital phase, are 
known up to 3.5PN order as far as orbital effects are concerned, and up to 
2.5PN and 2PN level for respectively ${\mathbf{S_{1,2}}}{\mathbf L}$ and
$\mathbf{S_1}\mathbf{S_2}$, $\mathbf{S_1}\mathbf{S_1}$,
$\mathbf{S_2}\mathbf{S_2}$ interactions.

According to studies in the non-spinning case 
\cite{Buonanno:2006ui,Baker:2006ha,Boyle:2007ft},
the TaylorT4 appears to be a very good approximant up to a frequency 
$\bar \omega = \pi\bar f_{GW}\simeq 0.1/M$ for the equal mass case, 
even though its faithfulness seems to worsen for different mass-ratios.

The PN evolution is halted when $\omega_{orb}$ reaches the value $\omm$ that is 
determined by comparison with numerical waveforms. For $\omega_{orb}>\omm$
($\omega_{orb}$ is monotonic here) the angular frequency is evolved according 
to 
\be
\label{eq:omfit}
\omega_{orb}(t)=\frac{\omega_1}{1-t/T_A}+\omega_0,\qquad 
\omm<\omega_{orb}\ {\rm and}\  t<t_{rac}\,,
\ee
where the three unknown parameters $\omega_{0,1}$ and $T_A$
are fixed by requiring that $\omega_{orb},\dot\omega_{orb}$ and $\ddot\omega_{orb}$
be continuous at the matching point defined by $\omm$, and $t_{rac}$ will be 
defined shortly.
In this phenomenological phase the amplitude of the waveform is 
evolved according to the lowest order formula eq.(\ref{eq:h22}), with $\iota$ 
and $\alpha$ frozen to their values at the instant of time when 
$\omega_{orb}=\omm$ (and keeping as usual $v=(M\omega_{orb})^{1/3}$)
\footnote{Clearly the approximation of constant $\iota$ and $\alpha$ in the 
phenomenological phase is a brutal one and it can be badly wrong in the case 
of strongly precessing systems (like systems with $\nu\ll 1$ where the 
more massive body has (nearly-)extremal spin not-aligned with the orbital 
angular momentum. However in those cases a phenomenological formula in the same
spirit of eq.~(\ref{eq:omfit}) can be written for $\iota$ and $\alpha$.}.\\

After this phenomenological phase, in order to smoothen the evolution of 
$\omega_{orb}$, which should settle at the ring-down value $\omrd$, it turned 
out useful to join the time evolution of eq.(\ref{eq:omfit}) with a smoothing 
function
\be
\label{eq:omsmooth}
\omega_{orb}(t)=\omrd-\pa{1-\frac{t}{T_A}}^2\omega_2\qquad 
t_{rac}<t\ {\rm and}\ \omega_{orb}<\omega_{mrd}\equiv 0.8\,\omrd
\ee
where $t_{rac},\omega_2$ are fixed by requiring continuity of $\omega_{orb}$
and $\dot\omega_{orb}$.
For $\omega_{orb}>\omega_{mrd}=0.8\,\omrd$ the waveforms is described by the 
damped exponential
\be
\label{ringdown}
\ba{rcl}
h^{(rd)}_{2,2}&=&\ds e^{-t/\tau}\paq{A\cos (\omrd t)+B\sin(\omrd t)}\,,\\
%h^{(rd)}_{2,-2}&=&\ds e^{-t/\tau}\paq{-B\cos (\omrd t)+A\sin(\omrd t)}\,,
\ea
\ee
where $A$ and $B$ are constant to be determined requiring the continuity of 
$h_{2,2}$.
For each multipolar mode defined by a pair $l,m$ there is an infinity of 
overtones with increasing damping factor, but for our practical purposes 
retaining only the first overtone is enough.\\
Note that the degree of continuity of the waveform across the phenomenological
-ring-down phase transition is not affected by changing the value of
$\omega_{mrd}$, but only by the number of overtones admitted to describe
the ring-down phase in eq.~(\ref{ringdown}), as it happens in the EOBNR
construction \cite{Buonanno:2007pf}.

The values of the ring-down frequencies and damping factors of
the three lowest overtones of the $l\leq 4$ modes can be read from 
\cite{Berti:2005ys} as a function of the mass and spin of the final object created 
by the merger of the binary system.
The final mass is determined by the algebraic sum of the constituents' masses
and the negative binding energy once $\omm$ is reached, and the final spin
according to the phenomenological formula given in \cite{Barausse:2009uz}.\\
The damped exponential is attached when $\omega_{orb}$ reaches 80\% of the $\omrd$
value as it turned out that this specific value allowed the best overlap 
between analytical and numerical waveform, irrespectively of the initial spins. 

The analytical waveforms just described has been quantitatively confronted with 
the numerically generated ones by computing the overlap integral 
\be
I_{h_1,h_2}\equiv \int h_1(f)h_2^*(f)\, df\,,
\ee
maximized over initial phase and time of arrival.
The angular frequency $\omm$ has been determined by a first set of numerically
produced short waveforms (4-6 cycles long) by picking the value which allowed a 
maximum overlap integral with the phenomenological waveform, with a 
determination precision of $\pm 10^{-4}/M$.

The numerical waveforms used in the present work have been generated with 
\texttt{MayaKranc}.  The grid structure for each run consisted of 10 levels of 
refinement provided by \texttt{CARPET} \cite{Schnetter-etal-03b},
a mesh refinement package for \texttt{CACTUS} \cite{cactus-web}.
Sixth-order spatial finite differencing was used with the BSSN equations 
implemented with \texttt{Kranc}~\cite{Husa:2004ip}.  The outer boundaries are 
located at $317M$ and the finest resolution is $M/77$.  Waveforms were 
extracted at $75M$.
A few waveforms were generated at resolutions of
$\{M/64, M/77, M/90\}$, and convergence consistent with our fourth order
code is found.  The short (long) runs showed a phase error on the order of 
$5\cdot 10^{-3}$ ($5\cdot 10^{-2}$) radians and an amplitude error of 
$\approx 2\%$ ($\approx 5\%$). We expect similar accuracy in all runs performed.

The numerical waveforms describe the $l=2$, $m=2$ mode and consist of two 
sets, both with equal mass
and spin magnitudes $|\mathbf{S_1}|=|\mathbf{S_2}|=0.6 m_1^2$.
The first set consisted in 24 few-cycle-long waveforms, representing mostly 
the merger and ring-down phases of a coalescence. They have been used to fix 
the values of $\omm$ for the corresponding values of initial spins. 
One of the binary constituent had initial spin ${\mathbf S_2}/m_1^2=(-0.6,0,0)$ 
in the reference frame in which the initial $\mathbf{\hat L}=(0,0,1)$. The 
different 
values of the first dimension-less spin have been obtained by rotating the 
$(0,0,0.6)$ vector by 15 degrees in the x-z plane. 
Once determined the values of $\omm$ for each waveform, the value of the 
matching frequency $\omm$ for generic spins has been determined by
assuming an analytical dependence of $\omm$ on ${\mathbf S_{1,2}}$, according to
\renewcommand{\arraystretch}{1.4}
\be
\label{eq:omegamatch}
\ba{rcl}
M\omm &=& a_0 + a_1 (S_{1z}+S_{2z}) + a_2 \delta (S_{1z}-S_{2z}) +
a_3 (S_1S_2)_\perp+a_4({S_1^2}_\perp +{S_2^2}_\perp) +\\ 
& & a_5 \delta ({S_1^2}_\perp -{S_2^2}_\perp) + a_6 (S_{1z}^2+S_{2z}^2) + 
a_7 (S_{1z}S_{2z}) +a_8\delta (S^2_{1z}-S_{2z}^2)+\ldots\,,
\ea
\ee
\renewcommand{\arraystretch}{1}
where the suffix $\perp$ stands for projection onto the plane perpendicular 
to $\mathbf{\hat L}$
and higher power of the spin variables have been neglected. The spin components
are understood in a frame where the orbital angular momentum is along the 
z-axis and their values are time-dependent, thus inducing
a mild time dependence in $\omm$. Note that because of the dependence
of the $\omega_{orb}$ evolution equation on $\mathbf{L}$, the spin components
parallel to the orbital angular momentum enter already at linear level, whereas
the dependence on the spin components in the plane of the orbit starts from 
the quadratic level. 
The $a_i$ coefficients may depend on the symmetric mass ratio $\nu$, but it is 
assumed here that they can be analytically expanded around their value at 
$\delta\equiv \sqrt{1-4\nu}=(m_1-m_2)/(m_1+m_2)=0$, according to
\be
\label{eq:expansion}
a_i(\delta)=a_i+\delta\, a_i^{(1)}+\delta^2 a_i^{(2)}+\ldots\,.
\ee
Since the set of simulations analyzed in this work do not include the case 
of unequal masses, it has been possible to fit only the first term in the 
expansion (\ref{eq:expansion}) for all the coefficients.
\footnote{Note that from the explicit formula for $\omega_{orb}$, see e.g. 
\cite{Buonanno:2009zt}, symmetric combinations of the spin components can only depend analytically on the 
symmetric mass ratio $\nu$, and thus will depend on even power of $\delta$, 
whereas anti-symmetric combinations do not appear for $\delta=0$.}

Given the specifics of the simulations we used (all having $\delta=0$ and
$\mathbf{S_2}\cdot\mathbf{\hat L}=0$ initially), it has been possible to 
determine only the coefficients $a_0,a_1,a_3,a_4,a_6$. These values have then been tested against
a second set of numerical waveforms, consisting of
8 long waveforms (12-15 cycles long), where all the three phases 
(inspiral, merger and 
ring-down) play a role. In particular in the long waveform case, even with the
initial condition ${\mathbf S_2}\cdot{\mathbf{\hat L}}=0$, the dynamical 
evolution has lead to generic ${\mathbf S_2}$ at the time of the onset of the 
phenomenological phase. Thus the determination of the unknown coefficients 
$a_{2,7,8,\ldots}$ will be necessary for an accurate estimation of $\omm$. 
Additional numerical simulations are required to obtain these extra information
and thus improve our model.\\
The result of the comparison between analytical and numerical waveform is the 
subject of the next section.

\section{Results}
\label{se:result}

\begin{center}
  \begin{figure}
    \includegraphics[width=.7\linewidth]{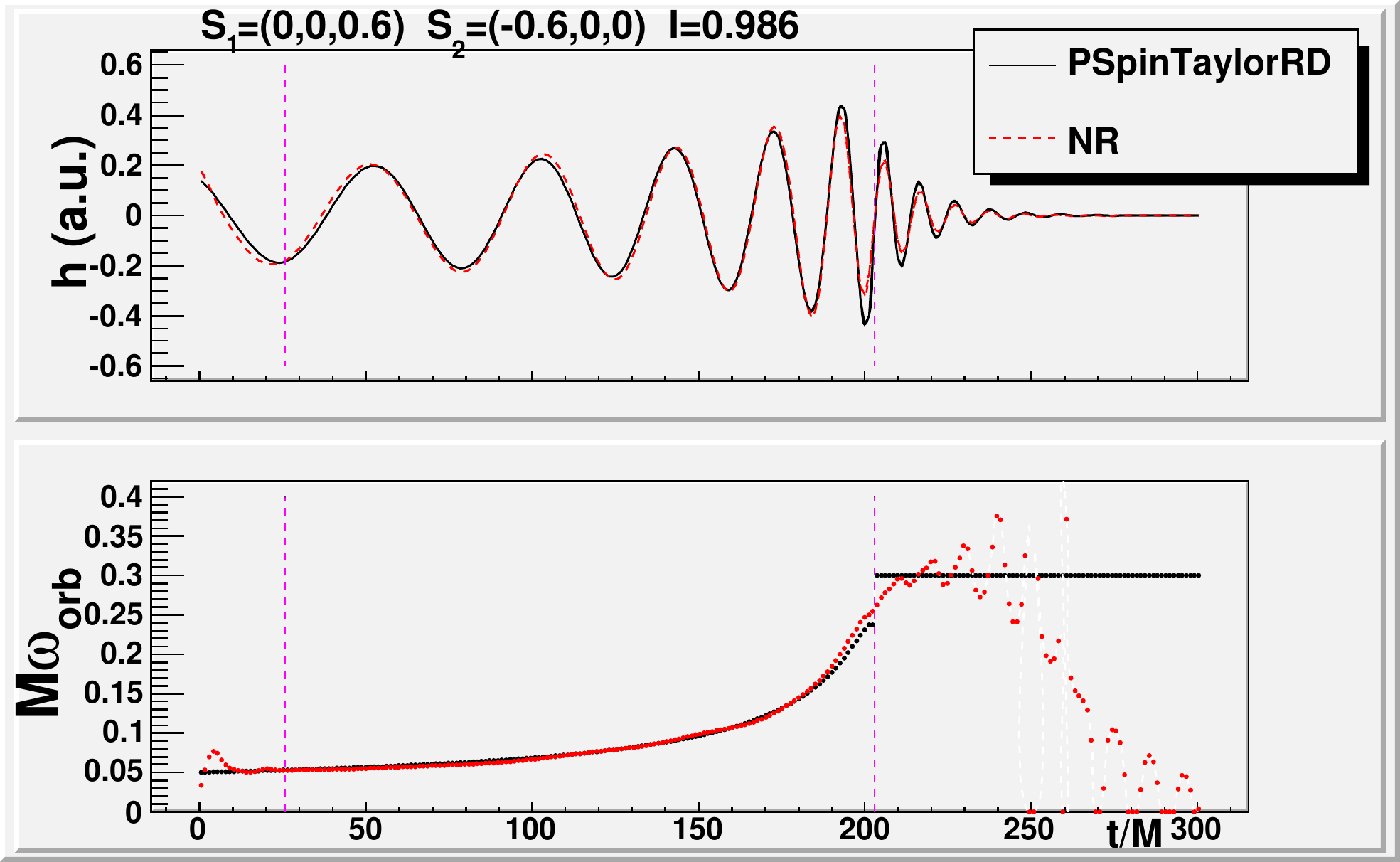}
    \caption{Comparison between the $l=2,m=2$ mode of analytical 
      (black, solid) and numerical (red, dashed) equal mass simulations: 
      time-domain waveforms are shown in the upper frame, the lower frame 
      shows $\omega_{orb}$ (given respectively by its PN value, by 
      eq.(\ref{eq:omfit}) and by $\omega_{rd}$ in the three different phases)  
      vs. the instantaneous frequency of the numerical waveform (computed as
      ${\rm Im}(\dot h_++i\dot h_\times)/|h_++ih_\times|$, being $h_+$ 
      ($h_\times$) the plus (cross) polarization of the numerical waveform).
      Initial spin configuration: $\mathbf{S_1}/m_1^2=(0,0,0.6)$, 
      $\mathbf{S_2}/m_2^2=(-0.6,0,0)$. Vertical dashed lines mark 
      the onset of the phenomenological phase parametrized by eq.~(\ref{eq:omfit})
      and the ring down phase parametrized by eq.~(\ref{ringdown}).
      The resulting overlap integral is 0.986.}
    \label{fi:over000}
  \end{figure}
\end{center}

\begin{center}
  \begin{figure}
    \includegraphics[width=\linewidth]{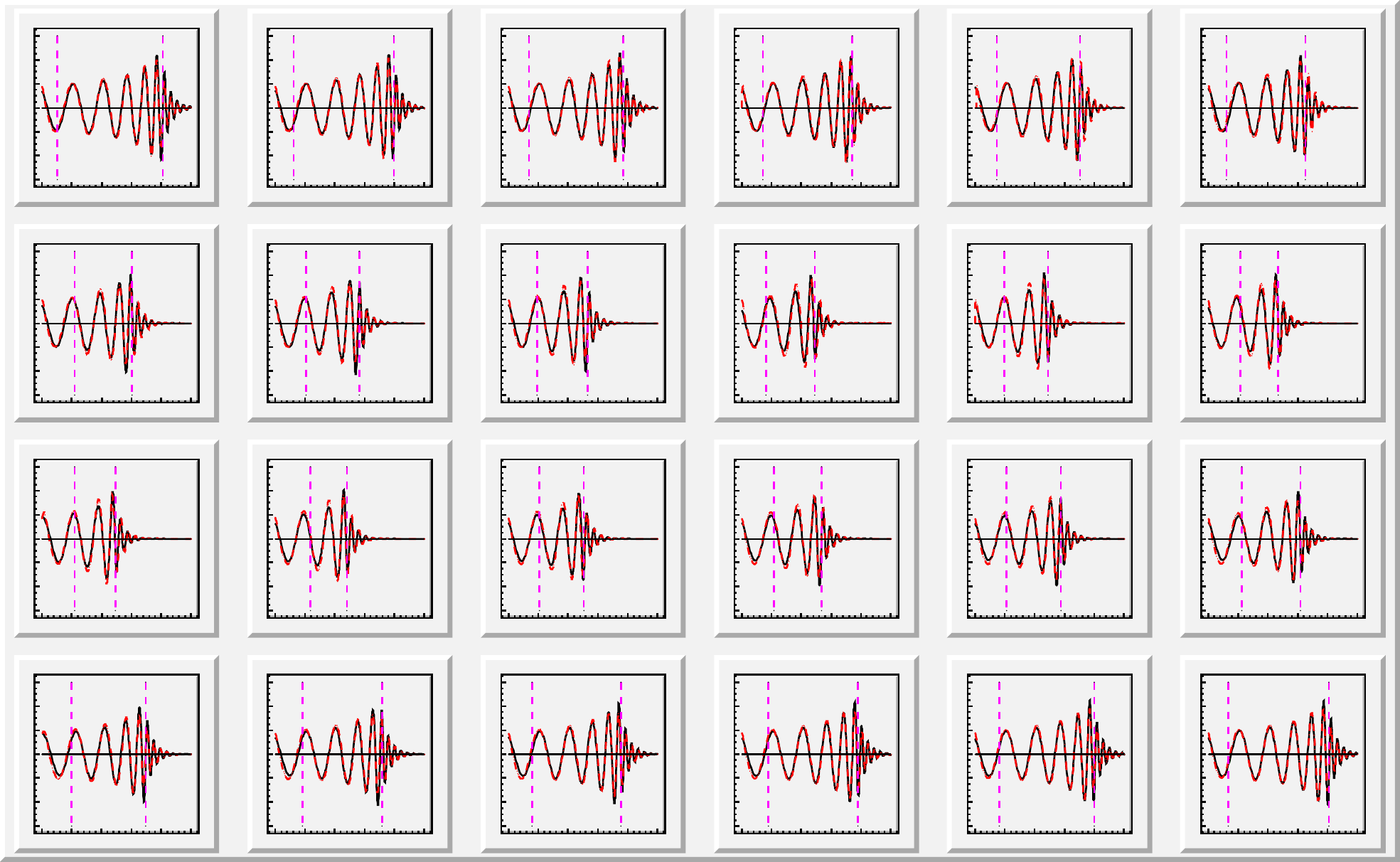}
    \caption{Summary of the comparison between the $l=2,m=2$ mode of analytical 
      (black, solid) and numerical 
      (red, dashed) waveforms. Initial spin configuration: $\mathbf{S_2}/m_2^2=(-0.6,0,0)$,
      $\mathbf{S_1}$ lies in the x-z plane. A rotation of $\mathbf{S_1}$ by 15 degrees 
      in the x-z plane takes from one plot to the following one, the first plot is the same as 
      in fig.~\ref{fi:over000}. Vertical dashed lines, x- and y-axis are the same as in 
      fig.~\ref{fi:over000}.}
    \label{fi:overshortall}
  \end{figure}
  \end{center}

The analytical waveforms have been calibrated by comparing them
with 24 short numerical simulations, with the result qualitatively
shown in figs.~\ref{fi:over000},\ref{fi:overshortall} and quantitatively 
reported in tab.~\ref{tab:short}.

The determination of the $\omm$'s giving the best overlap for different spin 
values allowed to evaluate some of the coefficients in the phenomenological 
formula (\ref{eq:omegamatch}), as given in tab.~\ref{tab:ais}.\\
The coefficients $a_{2,5,7,8}$ cannot be determined by the analysis 
performed because the terms they multiply vanish identically for the 24 
simulations considered. The $a_0$ coefficient has been determined by 
comparison with a non-spinning waveform\footnote{Note that given the 
moderate value of the spin magnitude, the plane of the orbit relative 
to these waveforms are only ``mildly'' precessing.}.

\begin{center}
  \begin{figure}
    \includegraphics[width=\linewidth]{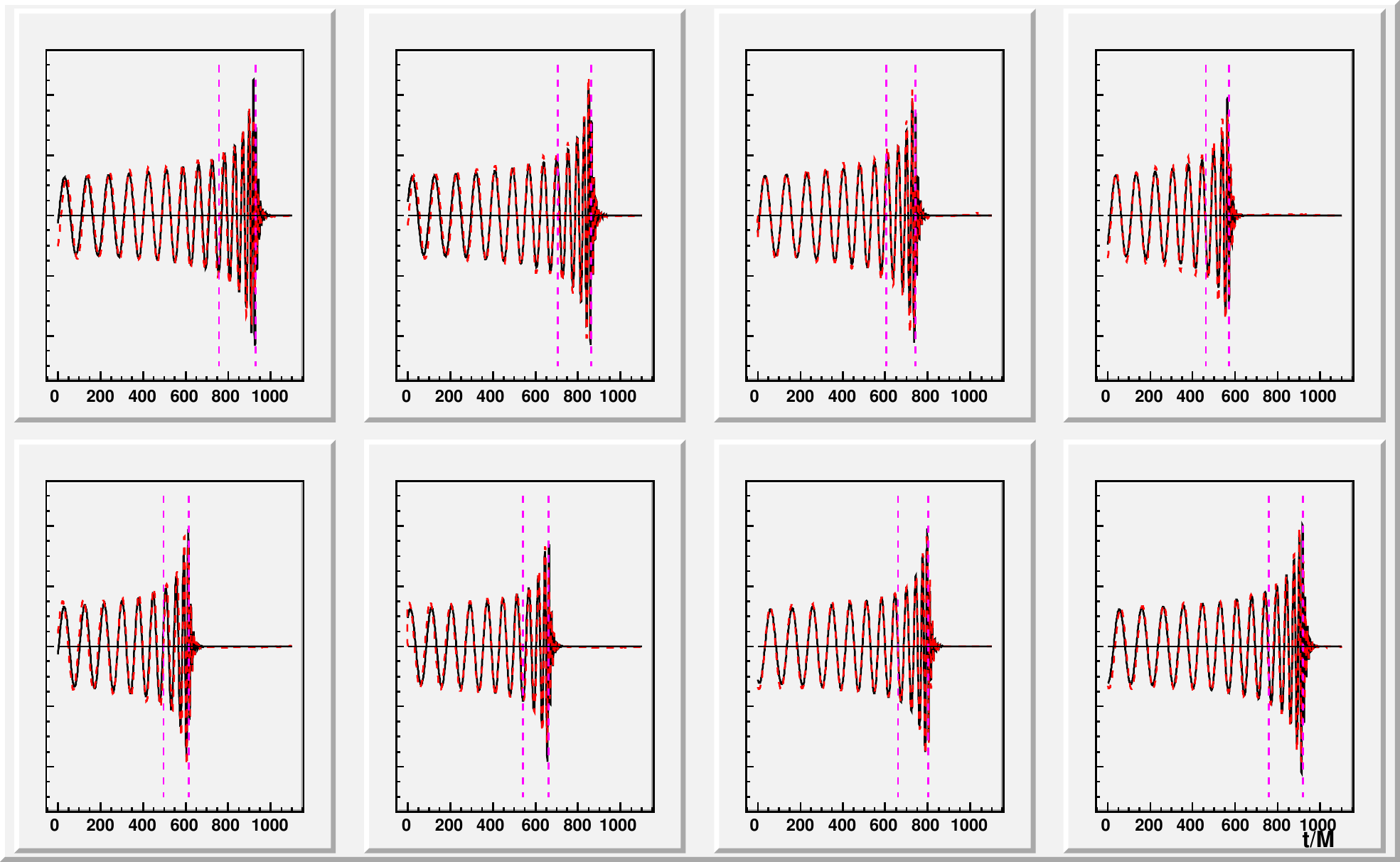}
    \caption{Comparison between the $l=2,m=2$ mode of analytical and long 
      numerical waveforms ($m_1=m_2$). Initial spin configuration: 
      $\mathbf{S_2}/m_2^2=(-0.6,0,0)$, $\mathbf{S_1}$ lies in the x-z plane. 
      A rotation of $\mathbf{S_1}$ by 45 degrees in the x-z plane moves the 
      initial condition of one plot to the following one, the first plot refers to the initial condition
    $\mathbf{S_1}/m_1^2=(0,0,0.6)$. Vertical dashed lines mark the onset of the 
      phenomenological phase and the ring-down phase, 
      x-axis is time in units of M.}
    \label{fig:long}
  \end{figure}
\end{center}

Having fixed the value of $\omm$ with some generality, it is possible to 
generate analytical waveforms with any specific initial condition without 
tuning any single parameter: the value of $\omm$ will be determined 
analytically via eq.~(\ref{eq:omegamatch}) with the unknown coefficients 
arbitrarily set to zero.
It is then possible to generate waveforms with no tunable parameters for 
comparison with the second set of long, numerically generated waveforms,
even though a wider range of initial conditions would be needed for
a more accurate determination of $\omm$ in the most generic case. 
Such comparison is summarized qualitatively in fig.~\ref{fig:long} and 
quantitatively in tab.~\ref{tab:long}.

 
\section{Conclusions}
\label{se:concl}
We presented an analytical method to produce 
complete gravitational waveforms from spinning coalescing binaries.
The free parameters of the model are the values of the orbital
frequency at the transition from the inspiral to the phenomenological phase
and from the phenomenological to the ring-down phase.
They have undergone a calibration process by confronting the analytical 
waveforms with numerically generated ones and in particular the accurate 
determination of the first of such parameters ($\omm$) turned to be crucial 
for a satisfactory waveform construction.\\
For the calibration process a first
set of (short) waveforms has been employed, obtaining overlap factors ranging 
between 0.97 and 0.99. Once the calibration has been obtained, the analytical 
waveforms, with no parameter to tune, had been confronted with a second set of
(long) numerical waveforms and overlap factors ranging from 0.95 to 0.99
have been obtained.
Better overlap is expected once
a larger number of available numerical waveforms makes it possible to have a 
more solid estimation of the matching frequency $\omm$
for generic spins.\\
The method illustrated here can be extended %straightforwardly used 
to produce other modes than the $l=2,m=2$
for which it has been tested. The $m=\pm 2$ quadrupolar modes are the only 
non-vanishing modes for head-on observation of a coalescing system, whereas to 
build physical waveforms for generic angle between the source and the observer 
all other modes are required, at least with $l\leq 6$.
In particular, as the quality of the numerical waveforms describing such modes 
improves, it would be interesting to verify if the calibration performed here 
on the dominant quadrupolar mode allow a good description of other modes.\\
The programs generating these phenomenological waveforms are written in the C 
language and are available from the LIGO Analysis Library (LAL) \cite{lal}.

\begin{center}
  \begin{table}
    \begin{tabular}{|c|c|c|c|}
      \hline
      $acos(S_1\hat L)[^o]$ & Overlap & $M\omm\times 10^{2}$ & $M\omrd\times 10^{2}$\\      
      \hline
      0  & 0.986 & 5.29 & 30.0\\
      \hline 
      15 & 0.991 & 5.37 & 29.5\\
      \hline
      30 & 0.991 & 5.44 & 28.8\\
      \hline
      45 & 0.986 & 5.49 & 29.0\\
      \hline
      60 & 0.977 & 5.55 & 28.0\\
      \hline 
      75 & 0.975 & 5.58 & 28.0\\
      \hline
      90 & 0.983 & 6.07 & 26.9\\
      \hline
      105 & 0.980 & 6.13 & 26.5\\
      \hline
      120 & 0.981 & 6.18 & 26.1\\
      \hline 
      135 & 0.987 & 6.00 & 24.7\\
      \hline
      150 & 0.983 & 6.36 & 26.4\\
      \hline
      165 & 0.984 & 6.59 & 26.0\\
      \hline
      180 & 0.985 & 6.47 & 26.3\\
      \hline 
      195 & 0.980 & 6.70 & 27.0\\
      \hline
      210 & 0.980 & 6.34 & 27.5\\
      \hline
      225 & 0.979 & 6.29 & 27.5\\
      \hline
      240 & 0.978 & 6.08 & 29.9\\
      \hline 
      255 & 0.977 & 6.03 & 28.9\\
      \hline
      270 & 0.971 & 5.68 & 29.4\\
      \hline
      285 & 0.977 & 5.62 & 29.6\\
      \hline
      300 & 0.980 & 5.48 & 30.4\\
      \hline 
      315 & 0.985 & 5.51 & 30.3\\
      \hline
      330 & 0.988 & 5.44 & 30.3\\
      \hline
      345 & 0.988 & 5.36 & 29.8\\
      \hline
    \end{tabular}
    \caption{Values of the overlap integral between the analytical and the 24
      short numerical waveforms. For reference, the values of $\omm$ 
      maximizing the overlap and of $\omrd$ are reported.}
    \label{tab:short}
  \end{table}
\end{center}
\begin{center}
  \begin{table}
    \begin{tabular}{|c|c|}
      \hline
      Coeff. & $\times 10^{-2}$\\
      \hline
      $a_0$ & 5.480\\
      \hline
      $a_1$ & -0.97\\
      \hline
      $a_3$ & 0.083\\
      \hline
      $a_4$ & 0.47\\
      \hline
      $a_6$ & 0.80\\
      \hline
    \end{tabular}
    \caption{Coefficients of eq.(\ref{eq:omegamatch}) as determined by comparison 
      of the analytical waveforms with 24 short numerical waveforms.}
    \label{tab:ais}
  \end{table}
\end{center}
\begin{center}
  \begin{table}
    \begin{tabular}{|c|c|c|c|}
      \hline
      $acos(S_1\hat L)[^o]$ & Overlap & $M\omm\times 10^{2}$ & $M\omrd\times 10^{2}$\\
      \hline
        0 & 0.9508 & 5.14 & 29.2\\
      \hline 
       45 & 0.9761 & 5.30 & 28.6\\
      \hline
       90 & 0.9935 & 5.50 & 27.2\\
      \hline
      135 & 0.9882 & 5.72 & 24.9\\
      \hline 
      180 & 0.9499 & 5.71 & 25.4\\
      \hline
      225 & 0.9605 & 5.70 & 25.9\\
      \hline
      270 & 0.9934 & 5.39 & 26.7\\
      \hline 
      315 & 0.9698 & 5.23 & 28.3\\
      \hline
    \end{tabular}
    \caption{Values of the overlap integral between the analytical and the 8
      long numerical waveforms. For comparison with the ``short'' case the 
      values of $\omm$ and $\omrd$ are reported.}
    \label{tab:long}
  \end{table}
\end{center}

\section*{Acknowledgments}
It is a pleasure to thank the organizers of the GWDAW-14 conference in Rome
for the stimulating scientific environment created by the meeting, as well
as Evan Ochsner for useful comments on the paper.  R.S. wishes
to thank the Physics Department of the University of Massachusetts at Amherst
for kind hospitality during the preparation of part of this work.
S.F. and L.C. wishes to thank the Universit\`a di Urbino for kind hospitality 
and the Istituto Nazionale di Fisica Nucleare (INFN) for support during the 
early stage of this work.
This work was supported  by the INFN and by the NSF grants PHY-0653550, 
PHY-0925345, PHY-0941417 and PHY-0903973.

\end{document}